\documentstyle[12pt]{article}
\begin{document}
\newcommand{\etal}{{\em et al\/}}

\begin{center}
{\Large\bf Determination of $^6$Li -- $^4$He interaction from 
multi-energy scattering data}

\vspace{20mm}

{\large R. S. Mackintosh and S.G. Cooper}%
\vskip0.5 cm%
{\large Physics Department, The Open University,\\ Milton Keynes
 MK7 6AA}
\vskip0.5 cm%
{\large V.I.Kukulin}
\vskip0.5 cm%
{\large Institute of Nuclear Physics, Moscow State University, 
Moscow 119899, Russia}

\vskip 0.5 cm%
r.mackintosh@open.ac.uk, s.g.cooper@open.ac.uk, kukulin@nucl-th.npi.msu.su%

\end{center}

\vspace{20mm}

\noindent {\bf Abstract:} We present the  first successful potential model 
description of $^6$Li -- $^4$He scattering. The differential cross-sections
for three energies and the vector analyzing powers for two energies were fitted
by a single potential with  energy dependent imaginary components.
An essential ingredient
is a set of Majorana terms in each component. The potential was determined
using a recently developed direct data-to-potential inversion method 
which is a generalisation of
the IP $S$-matrix-to-potential inversion algorithm. We discuss the problems
related to this phenomenological approach, and discuss the relationship
of our results to existing and future theories. 

\vspace{10 mm}
PACS: 25.55.Ci, 25.55.-e, 24.10.Ht

\vfill

\hfill\today
\pagebreak

\setlength{\parindent}{0.0 in}
\setlength{\parskip}{4 mm}

Various studies of the 
scattering of unpolarised~\cite{bingham} and polarised~\cite{green,george,rusek}
 $^6$Li from $^4$He using conventional 
potential models have failed to yield a 
fully satisfactory description of the data. Conventionally parameterised 
potentials fail 
seriously, even when the  absorption is allowed to be $l$-dependent.  
The  inclusion of cluster transfer processes 
within the higher order DWBA framework~\cite{green} and the 
inclusion of coupling to projectile excitations~\cite{rusek}, while clearly 
important, still do not result in satisfactory fits. 
In any case, the natural conclusion is still
that the detailed description of nuclear scattering in terms
of a potential model is not possible with $^6$Li -- $^4$He scattering. 
In this paper we suggest that this is not the case. 

Majorana terms are generally absent in published potential models
of $^6$Li -- $^4$He scattering. In a microscopic picture such as RGM 
these would arise, for example, from cluster exchange terms. In effect, this
is what the DWBA or CRC cluster exchange calculations include in a less
rigorous way. Green {\em at al}~\cite{green} show clearly how such
processes enhance the backward angle scattering.  It is well known 
that, within a potential model, Majorana terms lead to such enhancements.
 Determining the parameters of Majorana 
components is a non-trivial phenomenological problem since the radial
form is unknown. In fact, we know from potentials which exactly reproduce
RGM $S$-matrix elements for various systems 
that the radial forms of the Majorana terms will bear no simple relationship 
to those of the Wigner terms, see for example~\cite{np589592}.
Clearly, one must seek both Wigner and Majorana terms which are
unrestricted in form.  However,  model-independent fitting
methods bring to light deep ambiguity problems which make it a necessary 
(but not always sufficient) part of the fitting procedure 
to constrain the potentials to be 
reasonably smooth and depend  on energy in a physically reasonable way.  

In the present paper we apply a recently developed technique,
fulfilling the above requirements,  to analyse 
$^6$Li  elastic scattering  from $^4$He. 
We  have available numerical data for the following CM energies: 2.2 MeV and 7.85 
MeV~\cite{george}; 11.1 MeV and 15.0 MeV~\cite{green}. For the 15 MeV case
there is differential cross-section data only, but for all other cases
vector and tensor analysing powers have been measured, although in this 
initial study we only fit the 
vector analysing powers. Moreover, for reasons to be given,
the potentials we present here have not been fitted to the 2.2 MeV data. 

The data is fitted by
applying direct observable to potential inversion using a 
generalisation~\cite{kpz}
of the iterative perturbative, IP, $S$-matrix to potential inversion
method, a generalisation which has recently proven 
successful for the analysis
of proton scattering from $^{16}$O~\cite{npa618} and deuteron -- $^4$He
scattering~\cite{skm,puri}. The IP method for $S$-matrix to potential inversion
has been described many 
times~\cite{early,ketal1,candm,invprob}, 
so we briefly outline the
underlying concepts. The key idea is iteratively to correct a potential
 $V(r)$ by adding terms\begin{equation}
V(r) \rightarrow V(r) + \sum c_i v_i(r) \label{basic} \end{equation}
where $v_i(r)$ are members of a suitable set of `basis functions'
and $c_i$ are amplitudes derived from linear equations arising from the
 response, assumed linear, of the elastic scattering $S$-matrix to small
changes $\delta V$ in the potential:
\begin{equation}  \delta S_l = -\frac{{\rm i} m}{\hbar^2 k}
\int_0^{\infty} (u_l(r))^2 \delta V(r) {\rm d}r. \label{integ} \end{equation}
In Equation~\ref{integ}, the radial wavefunction for angular momentum
$l$ is normalised according to $u_l(r) \rightarrow I_l(r) - S_l O_l(r)$ where
$I_l$ and $O_l$ are the incoming and outgoing Coulomb wavefunctions. The 
notation
is simplified: $V(r)$ stands for real and imaginary, central and 
spin-orbit, Wigner and Majorana terms all of which can  be expanded 
in different bases; for the treatment of spin and multiple energies  
see the papers cited above.

The  generalised  IP method enables direct observable to 
potential inversion~\cite{kpz,npa618}.
At each iteration, the potential amplitudes $c_i$ in Equation~\ref{basic}
 are determined  by solving linear equations.
These arise from the minimisation of the
goodness of fit quantity $\chi^2$,
\begin{equation}
\frac{\partial \chi^2}{\partial c_i} 
=2\sum_{k,l}\left[\frac{\sigma_k-\sigma_k^{\rm in}}{(\Delta\sigma_k^{\rm in})^2}
\right]
\frac{\partial \sigma_k}{\partial S_l(E_k)}\frac{\partial S_l(E_k)}
{\partial c_i}
+2\sum_{n,k,l}\left[\frac{P_{kn}-P_{kn}^{\rm in}}{(\Delta P_{kn}^{\rm in})^2}
\right]
\frac{\partial P_{kn}}{\partial S_l(E_k)}\frac{\partial S_l(E_k)}
{\partial c_i}, \label{new}
\end{equation}
where $\sigma_k^{\rm in}$ and $P_{kn}^{\rm in}$ are the input experimental
values of cross sections and analyzing powers respectively ($n$ indexing
the spin related observables for spin 1 systems), and
\begin{equation}
\chi^2 = \sum^N_{k=1} \left(\frac{\sigma_k-\sigma_k^{\rm in}}
{\Delta \sigma_k^{\rm in}} \right)^2 +
\sum_n \sum^M_{k=1} \left(\frac{P_{kn}-P_{kn}^{\rm in}}
{\Delta P_{kn}^{\rm in}} \right)^2. \label{fourth}
\end{equation} Since we are 
fitting data for several, possible many, energies at once, 
the index $k$ indicates the energy 
as well as angle. For brevity, we shall  refer below to this method for 
observable to potential
inversion as the `Generalised Iterative Perturbative', GIP, method.
For recent applications to d + $^4$He scattering, 
including the  determination of phase shifts, refer to Refs~\cite{skm,puri,sk}.

An essential ingredient in the GIP approach is the `starting reference potential',
SRP, from which the iterative process starts. This is of great importance
in a system which is plagued by ambiguities even when fits are precise. It is the
main opportunity for the inclusion of {\em a priori\/} information derived
from physical insight. This is essential for systems such as that under 
discussion, for which there is no 
possibility of true model independent fitting of the kind associated, for
example, with electron scattering, although we see no reason not to
aspire to fits of such quality. 

We seek eight potential components in all: real and imaginary central 
and vector spin orbit
terms, each a sum of Wigner and Majorana terms:
$V_{\rm W}(r) + (-1)^l V_{\rm M}(r)$.
For the real potential, the inversion determines
the radial shape of the potential and the coefficients of a polynomial expansion
in energy which multiplies the radial form. That is, the algorithm
determines $V(r)$ and $\xi_i$ in the expression:\begin{equation}
{\rm Re}V(r,E) = V(r)(1 + \xi_1 E + \xi_2 E^2 + \ldots).\label{eq:VrE} \end{equation}
(More general forms are allowed by the algorithm but are not relevant to this work.)
The imaginary potential is also parameterised as the product of a radial form to
be determined and a function of energy. The energy function 
takes the form of a leading power term plus a polynomial series which is
 not used in most cases described here. 
The form was determined by the requirement with certain cases of light 
nuclei that the imaginary part be zero below the reaction threshold, $E_0$. 
The leading term is:\begin{equation} {\rm Im}V(r,E) =
W(r) \left( \frac{E-E_0}{E_{\rm ref} -E_0}\right)^p \label{eq:imag} \end{equation}
so that in the present case, taking $E_0=0$ and absorbing the other constants
into the radial form, we have simply an $E^p$ dependence. The inversion 
procedure does not automatically 
adjust $p$. In the present case we have fixed  $p=1$ but there is no restriction 
in principle on $p$ to integral or positive values.

All the potentials we present were found using a Gaussian
inversion basis, generally starting with three or four terms for each 
component. There is a natural limit
to the desirable basis dimension since, as this is increased, there
comes a point where there
is a rather sudden onset of oscillations in the potential.
Only vector spin-orbit interactions were included
and no attempt was made to fit tensor analysing powers. A coupled channel
extension of IP inversion to determine tensor potentials is 
presently under development~\cite{ckpm}. Studies with deuteron
scattering suggest that, although omission of $T_{\rm R}$ terms might compromise
the vector interaction somewhat, the 
general features of the central potential should be little affected.

{\em Preliminary investigations\/.} Our initial calculations involved only
the two higher energies. We first used as SRP
various the potentials given by Green {\em et al\/}~\cite{green}. In this
way we achieved our best overall fit to the 11.1 MeV and 15 MeV data. 
However, we discount this potential as
a possible physical solution since the volume integral per nucleon pair
for the real central Wigner component,
$J_{\rm R}$, was about 800 MeV fm$^3$, much higher than the volume integral that
might be expected from  systematics. Indeed,  $^6$Li scattering phenomenology
suggests a potential which is weaker in the surface,  
as a result of breakup processes, than the M3Y folding model~\cite{m3y} 
potential, see below. It seems likely that 
 Green {\em et al\/}~\cite{green} found potentials with very large $J_{\rm R}$ as
a consequence of attempting to fit aspects of
the data which cannot be fitted without Majorana components. 
Of the various terms of the SRP,
it is the choice of the real central Wigner term, by far the largest component,
which most influences the final potential.

{\em Subsequent investigations\/.}
For the fits that we present here  the central Wigner component of the SRP was 
the M3Y density independent folding model~\cite{m3y} potential incorporating
a $^6$Li density of Suelze {\em et al}~\cite{suelze}.
The imaginary and spin-orbit components were taken from Green {\em et al\/}.
 We first found a potential which simultaneously fitted  the $E_{\rm CM} =$
11.1 and 15.0 MeV data, and then found a simultaneous fit to the $E_{\rm CM} =$
7.85, 11.1 and 15 MeV data.
In the following figures and text, these data are referred to by the projectile 
($^6$Li) energies, 19.6, 27.7 and 37.5 MeV, respectively. 
We comment below on the 5.5 MeV ($E_{\rm CM} = 2.2$ MeV) data.

First, we fitted the  27.7 and 37.5 MeV data from 
FSU with a potential for which the real Wigner central component had a volume integral
$J_{\rm R} = 339$MeV fm$^3$, i.e.\ somewhat below that for the M3Y folding model,
but still   much closer to it than the initial potentials with 
$J_{\rm R} = 800$ MeV fm$^3$~\cite{green}. This potential, which we refer to as
2EN-1, is included for comparison
in Figures 3 -- 5 to be referred to below. It had no pathologies such as
unitarity breaking in particular $lj$ partial waves. This potential did not give a 
satisfactory fit to the Wisconsin~\cite{george} data. 

An attempt to fit the two Wisconsin data sets together was not very successful,
and it seemed that the very low energy, 5.5 MeV, data were too 
easily fitted. 
The backward angle oscillations in the other cases are evidence of two amplitudes
interfering, and this requires Majorana terms in a single particle model. However,
there are other ways of getting the structure-less backward angle rise at 
5.5 MeV, and
this seems to have undermined the fitting process.

In Figures 1, 2  and 3, 
we present a simultaneous fit to the 19.6, 27.7 and 37.5 MeV
data. The most serious mismatch is to the forward angle 
37.5 MeV data although this is less serious than for the
`two-energy' fit. A suggestion of such a forward angle mismatch is present
in an existing fit~\cite{green} to the 37.5 MeV  data, and it was decided
not to pursue perfect fits at these angles.
 The imaginary terms were energy dependent (except the Wigner SO term)
as described above with $p=1$, $E_0=0$ and $E_{\rm ref}= 19.6$.
The real terms (not spin-orbit) turned out to be very
weakly energy dependent, $\xi_1$ being very small. The  potential corresponding to
Figures 1 to 3 is referred to as 3EN-1 in Figures 4 and 5 which also show
the `two-energy potential' labelled 2EN-1, 
and an alternative potential, labelled 3EN-2. The real, central, Wigner
component of the adopted potential 3EN-1 has volume integral $J_{\rm R} = 411$
MeV fm$^3$ and rms radius 3.042 fm.

The potential 3EN-2 is more oscillatory than the others,
and the fit is markedly poorer, especially for  19.6 MeV. It
was found in an attempt to eliminate  unitarity breaking
by 3EN-1 for $l=2$ and $j=3$ at 19.6 MeV. Although  2EN-1
did not show such a pathology,  it is, on balance, a less reasonable
representation of the data. With discrete two-step data fitting,
i.e., data $\rightarrow S_{lj}$ followed by  $S_{lj} \rightarrow V(r)$
inversion, it should be possible to constrain $|S|\le 1$;
the next stage in the development of the present method will
be to incorporate such a constraint.

Unitarity breaking is thus a symptom of the difficulties presented by these data,
presumably due to strong non-localities and channel couplings.
The ambiguity problems were greater than
presented by d + $^4$He data,
although these were significant~\cite{skm,puri}. 
The question of what we can say reliably about the
potential is not straightforward. Should we completely discard a potential
which breaks unitarity in one $lj$ channel? In the present case $S_{23}= 1.26$,
unquestionably an unphysical feature. However, we argue that
the key properties of the potential are determined. The problem is
that, because one must carefully limit the
dimension of the inversion basis, it is unlikely  that a potential
which does precisely fit the data will lie within the space spanned by the
particular basis employed. It appears that  potentials which are
found to give reasonable fits
may, among their many components, have an emissive region  
which happens to coincide
with a substantial value of $|\psi|^2$ for some particular $lj$
leading to $|S|>1$ for that $lj$.  Many inversion studies with RGM
$S$-matrix elements have shown that emissive regions certainly do occur
in local potentials representing the RGM $S$-matrix (with all $|S|\le 1$,
however) as a result of non-locality arising from exchange and channel
coupling effects. 

Whether or not this picture can be substantiated, 
we have shown that there exists a 
potential, having reasonable central Wigner real and imaginary 
terms and also substantial
Majorana terms, which does give {\em much\/} better fits than other potential models.
The fact that there also exists a  potential which gives a  better fit
to just the 27.7 and 37.5 MeV data but which is certainly unphysical 
(having perhaps twice the expected
$J_{\rm R}$) shows two things: (i) the ambiguity problem is deep and treacherous, and 
(ii) the OM fits in the original papers found unreasonably deep
potentials because they were trying to fit the backward angles without Majorana terms.
This last point implies that, if we have shown nothing else, we have shown that
$^6$Li + $^4$He scattering is compatible with the expectations of folding models
(as long as some account of exchange via Majorana terms or otherwise is included.)
To emphasise the reasonableness of the potential, we compare in Figure 6 the
real central Wigner terms for the 3EN-1 and 2EN-1 potentials
with the M3Y density independent folding model potential incorporating
a $^6$Li density of Suelze {\em et al}~\cite{suelze}. It is clear that
the three energy potential 3EN-1 is remarkably close to
the folding model potential except in the surface;
just where one expects repulsive effects due to the breakup of $^6$Li,
see Ref.~\cite{early} and papers cited therein. The alternative folding
model\footnote{Based  not on an effective  NN-force like M3Y but 
on well established   cluster-cluster 
potentials, correctly fitted to the data.}
 potential shown in Figure 8 of Kamal {\em et al\/}~\cite{kametal} is just
a little deeper than 3EN-1 but remarkably similar in shape to both
it and the M3Y potential.

One  conclusion of this work is that it is {\em essential\/} in this sort of
fitting to include {\em a priori\/} knowledge, in this case by employing
an M3Y based folding potential as starting potential (SRP). Two complementary
sources of such information are: (i) RGM that includes exchange exactly and,
 by way of inversion of 
RGM $S_{lj}$, that yields useful starting forms for central and vector
spin-orbit Wigner and Majorana terms; (ii)  $\alpha +d+\alpha $ cluster models
with $d$-exchange, which can include the D-state components which are omitted
in current RGM calculations. These latter will be useful
at the next step of this research programme, i.e. when tensor interactions are
included and tensor analysing powers are fitted. How might 
our analysis be affected by our omission of
tensor degrees of freedom? We explored this
with test cases involving  deuterons with
$S_{lj}$ calculated with potentials which included a strong $T_{\rm R}$ interaction.
Inversion of these  $S_{lj}$ gave
potentials with just central  and vector spin-orbit interactions and also
Majorana components.  Though not a rigorous argument, 
this suggests that the central Wigner terms are little affected 
by the omission of tensor interactions from the analysis, but
that the  spin-orbit interactions and Majorana terms will be modified to some extent.
This corresponds to the fact that differential cross-sections
are hardly modified by tensor interactions, which do, however,
somewhat affect vector analysing powers.

We can now draw the following conclusions:
(i) it is possible to get a reasonable simultaneous fit to the scalar and 
vector observables
for three energies with a single energy dependent potential; 
(ii) the real, central, Wigner term
is very close to what would be expected from a folding model; (iii) substantial
Majorana terms must be included, even though
their specific nature might be obscured by the neglect of tensor forces;
(iv) one cannot assume the Majorana terms to be of the form $1 + \alpha (-1)^l$
times the real components. In cases where RGM or cluster exchange theories imply
the existence of Majorana terms, they must be included
in phenomenological analyses. If
they are omitted, one cannot achieve  the quality of fit
which is necessary for  determining even the Wigner components.
Thus,  the best fit pure Wigner potentials,
which did not fit the data very well,
had much larger $J_{\rm R}$ values than folding models would imply.
A significant limitation at present, and a reason for not pressing for
closer fits, is the assumption that the energy dependent potentials
are of fixed radial form.

The GIP method has much scope for development as a
powerful tool for analysing nuclear scattering. Elsewhere~\cite{skm},
we have described its utility for phase shift analysis
in the common situation where the data is incomplete.
The next developments will be the
introduction of unitarity constraints and tensor interactions. 
If successful, the quantitative evaluation of microscopic theories 
will become possible. This 
is of great interest for $\alpha + ^6$Li  for which
there are reasons to anticipate parity dependent 
tensor forces. 
 Such theories are very hard to test by conventional methods.
This is because, on the one hand, the ambiguity and fitting problems 
faced by conventional phenomenology are extreme, and, 
on the other hand, because the 
complexity of the theory for this ten body system is such that even
the most elaborate {\em ab initio\/} scattering calculations are 
likely to give modest fits to the data, making direct evaluation of the theory
problematic. This is certainly the case for the d + $^4$He six
nucleon\footnote{One problem is including the deuteron and alpha particle
 D-states within the fully antisymmetrized RGM formalism. It is the deuteron
D-state, of course, which
leads to the $T_{\rm R}$ interaction in the inter-nucleus potential.}  
scattering system~\cite{kpz,puri,sk,dalph}, and will be even more so here.
However, when the present method is applied with {\em a priori\/} 
information in the form of SRPs derived by inversion from theory,
and with the inversion potential constrained to depart as little as possible
from the SRP, then the final departure from the SRP required to get a 
perfect fit will furnish a quantitative evaluation of the theory. 

When it was first found long ago that, contrary to  expectations, 
nucleon scattering could
be described by a single particle model, it was considered remarkable. 
We have shown that the range of application of this  model is
by no means exhausted, but a necessary generalisation is that all
components must be free to be parity dependent. 
 
\section*{Acknowledgements} We are much indebted to K. Kemper of FSU and
E. George of Wisconsin for supplying us their impressive numerical data.
We are also grateful  to the Engineering and Physical Science Research Council
of the UK for grants supporting S.G. Cooper and to the Royal Society for supporting
a visit by Professor Kukulin.

\newpage
{\sc Figure Captions}

\setlength{\parindent}{0.0 in}
\setlength{\parskip}{0.3 in}

{\sc Figure 1.}  Fit to  $^6$Li -- $^4$He differential cross-section
and analysing power data, for incident $^6$Li energy 
19.6 MeV, with parity dependent and
energy dependent potential 3EN-1.
 
{\sc Figure 2.}  Fit to  $^6$Li -- $^4$He differential cross-section
and analysing power data, for incident $^6$Li energy 
27.7 MeV, with parity dependent and
energy dependent potential 3EN-1.
 
{\sc Figure 3.}  Fit to  $^6$Li -- $^4$He differential cross-section
 data, for incident $^6$Li energy 
37.5 MeV, with parity dependent and
energy dependent potential 3EN-1.
 
{\sc Figure 4.} Wigner components of potentials fitting $^6$Li -- $^4$He
scattering data. From the top, real and imaginary central, real and imaginary 
spin-orbit. The solid line is 2EN-1 fitting data for
the two higher energies; the dashed line is the best `three energy' potential, 
3EN-1, and the dotted line represents 3EN-2, an unsuccessful attempt to 
improve the fit, showing the tendency to become oscillatory. The imaginary 
terms are evaluated for $E=E_{\rm ref}$.

{\sc Figure 5.} Majorana components as for Figure 4.

{\sc Figure 6.} Comparing the real, central, Wigner component of the
potentials 3EN-1 and 2EN-1 with the M3Y folding model potential.
  
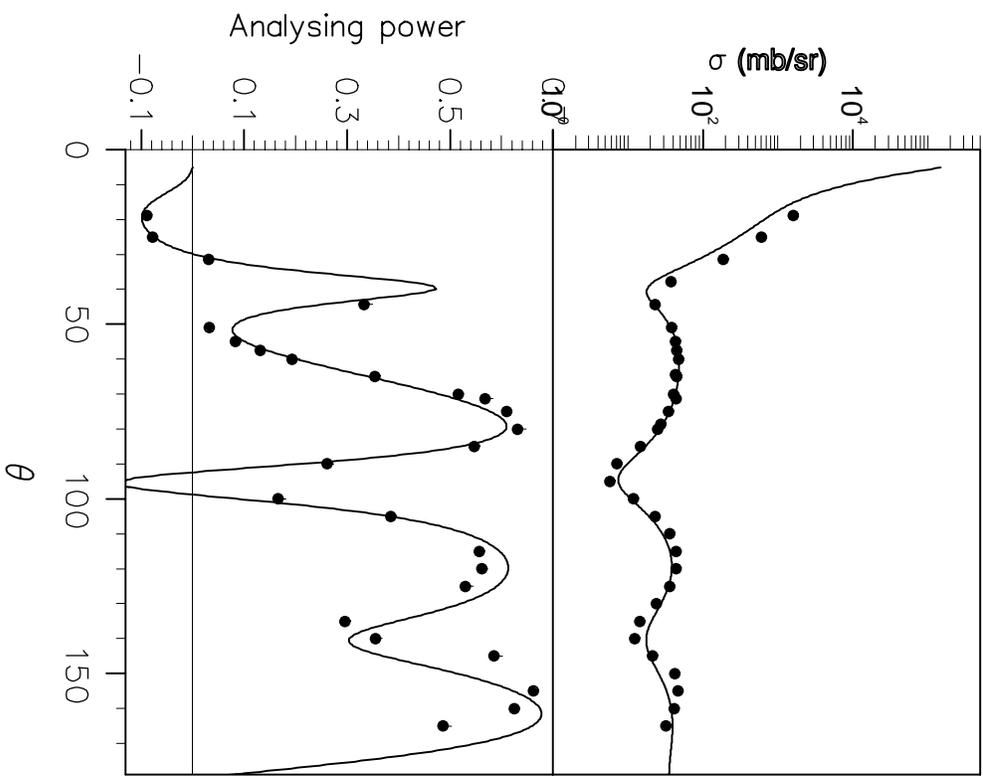
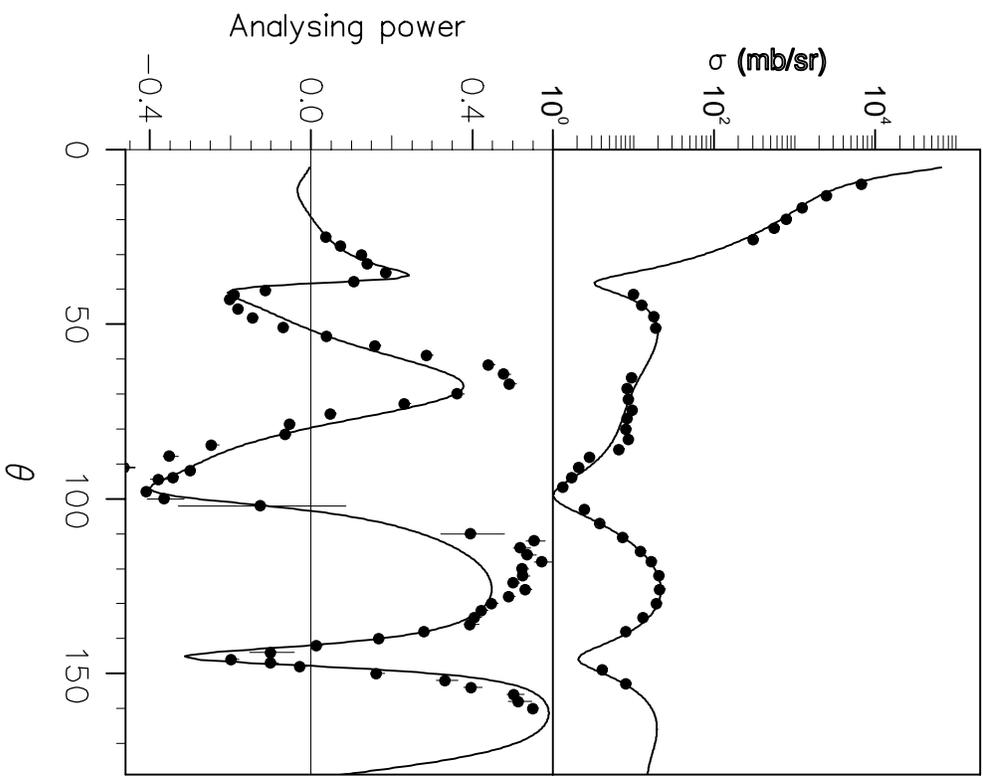
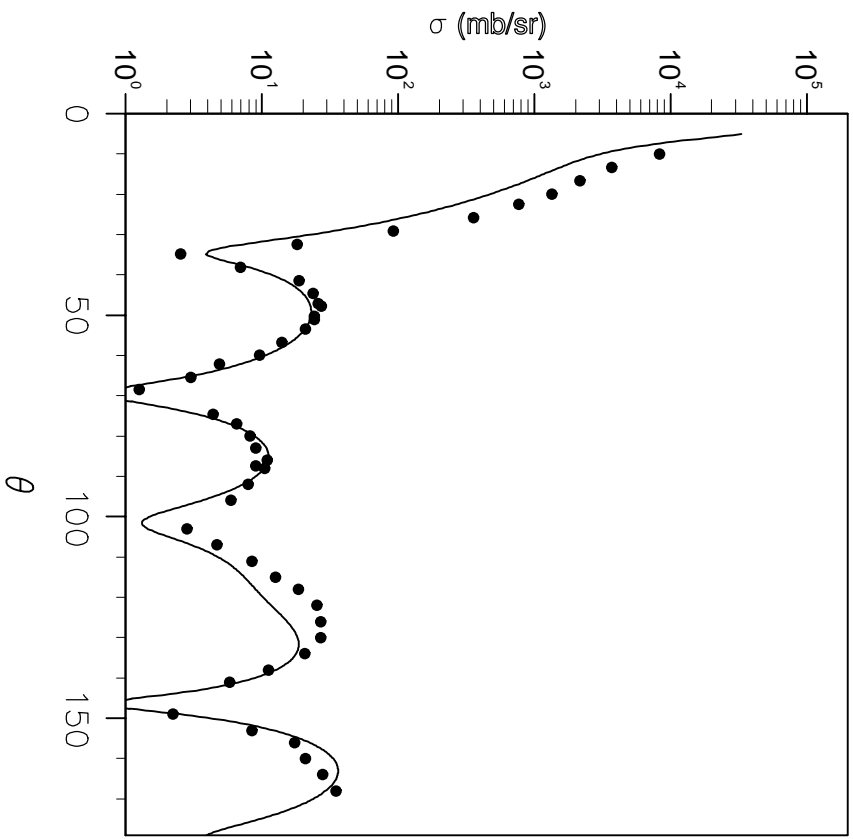
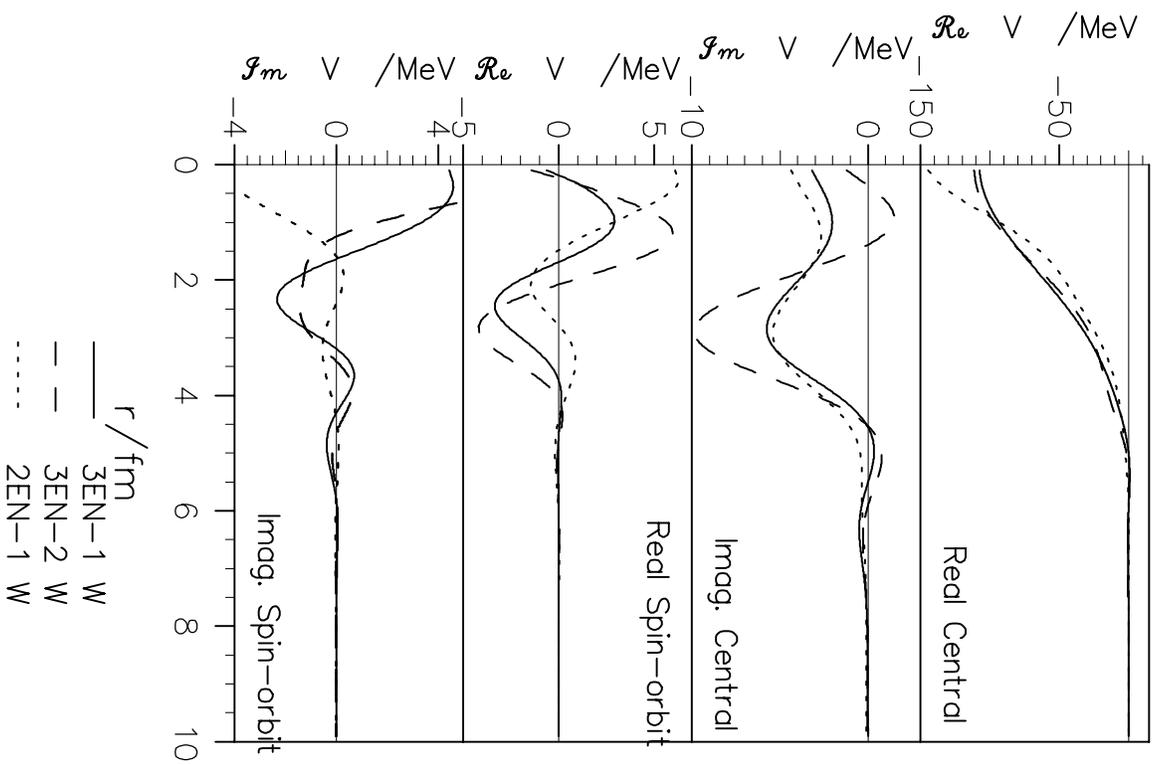
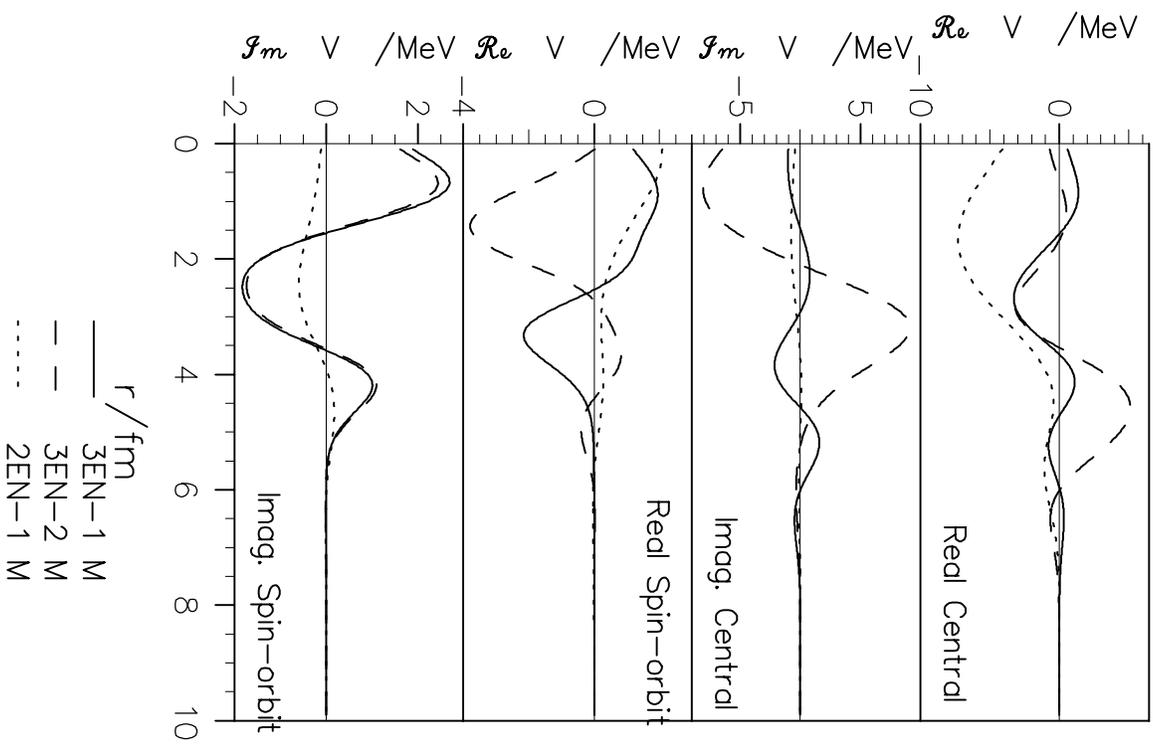
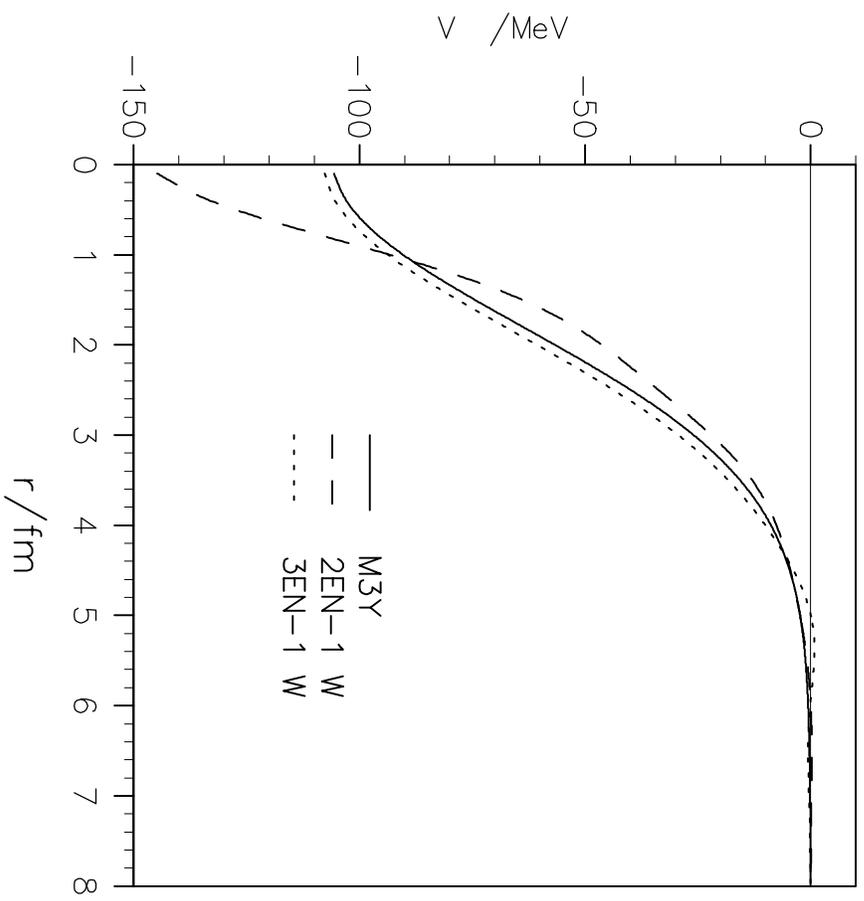
\end{document}